\documentclass[final,3p,8pt]{elsarticle}

\usepackage{amssymb}
\usepackage{array}
\usepackage{mathrsfs}
\usepackage{amscd}
\usepackage{amsmath}
\usepackage{amsfonts}
\usepackage{amsthm}
\usepackage{psfrag}
\usepackage{textcomp}
\usepackage{bm}
\usepackage{dsfont}
\usepackage{algorithm}
\usepackage{algpseudocode}

\usepackage{amsthm}

 \usepackage{lineno}

\journal{Wave Motion (Elsevier)}

\usepackage{amssymb}

\biboptions{square,sort&compress}

\usepackage[figuresright]{rotating}
\usepackage{moreverb}
\usepackage{amssymb}
\usepackage{array}
\usepackage{mathrsfs}
\usepackage{amscd}
\usepackage{amsmath}
\usepackage{amsfonts}
\usepackage{amsthm}
\usepackage{psfrag}
\usepackage{textcomp}
\usepackage{bm}
\usepackage{color}
\usepackage{float}
\usepackage{afterpage}
\usepackage{tabularx}
\usepackage{booktabs}
\usepackage{romannum}
\usepackage{multirow}
\usepackage[normalem]{ulem} 
\usepackage{cancel}

\begin{document}
	

\begin{frontmatter}





\title{Weak scattering formulation for flexural waves in thin elastic plates with point-like resonators}


\author[upv,icl]{Mario L\'azaro\corref{cor}}
\author[icl]{Marc Mart\'i-Sabat\'e}
\author[icl,icl2]{Richard V. Craster}
\author[upv]{Vicent Romero-Garc\'ia}

\cortext[cor]{Corresponding author. \newline $\mbox{ $\quad \,$}$  Email address: malana@upv.es }
\address[upv]{Instituto Universitario de Matem\'atica  Pura y Aplicada, 
Universitat Polit\`ecnica de Val\`encia, 46022 (Spain)}
\address[icl]{Department of Mathematics, Imperial College London, London, SW7 2AZ, UK}
\address[icl2]{Department of Mechanical Engineering, Imperial College London, London, SW7 2AZ, UK}

\begin{abstract}

A theoretical study on the weak scattering formulation for flexural waves in thin elastic plates loaded by point-like resonators is reported. Our approach employs the Born approximation and far-field asymptotics of the Green function to characterize multiple scattering effects in this system. The response of the system to an incident wave can be expressed as a power series expansion, where each term introduces higher-order scattering components. The behavior of this series is governed by the spectral properties of a specific matrix, which dictates the convergence and accuracy of the weak scattering approximation. By decomposing the scattering response into geometric and physical contributions, we establish a condition for weak scattering in terms of these two factors. This formulation provides a systematic framework for assessing the validity of low-order approximations and understanding wave interactions in complex media. Numerical examples illustrate the accuracy of the weak scattering condition in periodic and random distributions of scatterers, highlighting its implications for wave control and metamaterial design.

\end{abstract}


\end{frontmatter}


\section{Introduction}

Metamaterials are engineered composites with properties that surpass those of their individual components. The research in this field has been extensively active emerging in electromagnetism \cite{pendry2000negative} but quickly extended to other classical wave systems \cite{schurig2006metamaterial,valentine2008three,zhang2009focusing,zhang2004negative,chen2006active}. In the field of elasticity, elastic metamaterials have been developed in the last decade as an efficient platform for achieving wave control. The capability for manufacturing these structures at various scales has unveiled the potential of mechanical metamaterials for providing solutions for challenges such as vibration isolation \cite{alshaqaq2020graded,de2020graded}, energy harvesting \cite{badreddine2012broadband,miniaci2016large,achaoui2017clamped}, or cloaking \cite{farhat2009ultrabroadband,colquitt2014transformation}. \\

On the one hand, there is a wide variety of structures used as mechanical metamaterials, spanning from discrete mass-spring networks \cite{rosa2022small}, continuous two-dimensional phononic crystal \cite{torrent2013elastic,wang2015topological}, bio-inspired \cite{mazzotti2023bio} or non-periodic \cite{marti2021dipolar,marti2021edge} lattices. The achieved effects are numerous, such as local resonances and Bragg scattering \cite{achaoui2011experimental}, tailored dispersion \cite{miao2023deep}, non-linear behavior \cite{fronk2023elastic,haberman1976nonlinear} or topological effects \cite{carta2020chiral,wiltshaw2023analytical} among others. Most of these effects rely on the strong interaction between the constituents of the material, having a big impact on the capabilities of the heterogeneous medium to control wave propagation.\\

On the other hand, there is a huge amount of complex media where the constituents have a weak influence on the properties of the resulting material. In many body systems, such as complex structures with discrete resonators embedded in continuum systems, the multiple interaction between the elements can lead to multiple order of scattered waves. When the influence of the scatterers is small compared to the propagation of the incident wave in the medium, reduced models can be developed, encapsulating the necessary Physics to understand the problem in the desired scale. The Born approximation \cite{born1926quantenmechanik}, first used in the context of quantum physics, consists of neglecting these higher order scattered waves, considering that the wavefield arriving to the resonators in the problem is the same as the incident wavefield to the system, i.e., assuming weak scattering conditions. Born approximation has been widely applied in optics to understand propagation through random scattering media \cite{korotkova2015design,kvien1995validity} but has also been applied to elastic media \cite{gubernatis1977born,hudson1981use}. In the context of thin elastic plates in which only flexural waves can, it has been applied to cylindrical inhomogeneities in the plate \cite{lu2017scattering,rohde200712d,wang2005scattering}.\\

In this paper, we study the weak scattering approximation in the multiple scattering problem for a collection of point-like resonators  loaded to a thin elastic plate. Born approximation of the scattering solution and far-field asymptotics of the Green function will be used in order to separate the geometric (the effects coming from the spatial distribution of the resonators) and the physical (the effects coming from the local resonances) dependencies of our solution. This will allow us to establish a weak scattering condition in terms of these two quantities, giving Physical insight to this kind of many-body systems. \\

The article is organised as follows: Section \ref{sec:MultipleScatteringInThinElasticPlates} reviews the basic formulation for the multiple scattering of point-like resonators  loaded to a thin elastic plate. Section \ref{sec:WeaklyScatteringResonators} develops the weak scattering formulation for the problem in terms of Born approximation and far-field asymptotics of the Green function, and stating both the conditions and the bounds for the weak scattering assumption. Section \ref{sec:NumericalExamples} illustrates the weak scattering formulation and the relationship between the spectral radius and the geometrical distribution of resonators. We consider three numerical examples. A discussion about the relevance of geometry and physical properties is done, and the comparison between the weak scattering and the full multiple interaction is shown.

\section{Multiple scattering in thin elastic plates}
\label{sec:MultipleScatteringInThinElasticPlates}

\begin{figure}[h]%
	\begin{center}
		\begin{tabular}{c}
			\includegraphics[width=9cm]{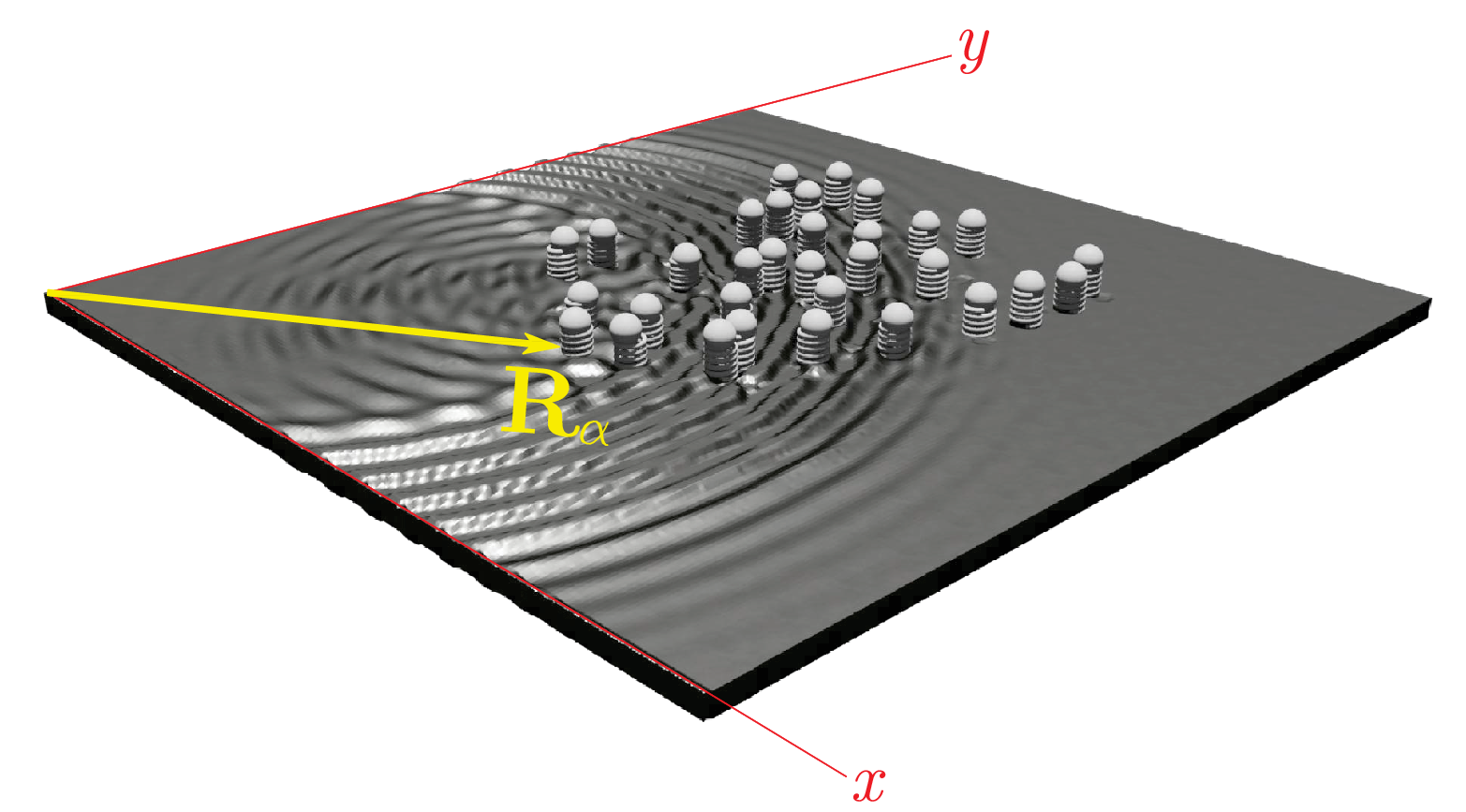} 
		\end{tabular}			
		\caption{Infinite plate with a distribution of $N$ resonators of $m_\alpha$ (mass) and $K_\alpha$ (stiffness) properties placed at positions $\mathbf{R}_\alpha$, for $1 \leq \alpha \leq N$.}%
		\label{fig01}%
	\end{center}
\end{figure}

We consider a thin elastic plate of thickness $h$, mass density $\varrho$, Young's modulus $E$ and Poisson ratio $\nu$. The propagation of flexural waves in the out-of-plane vertical displacement is given by Kirchoff-Love theory. After considering a time-harmonic regime $w(\mathbf{r},t) = w(\mathbf{r}) e^{-i \omega t}$, the equation of motion is given by \cite{Doyle-1997}
\begin{equation}
	\left(D \nabla^4 - \varrho h \, \omega^2 \right) w(\mathbf{r}) = q(\mathbf{r}),
	\label{eq001}
\end{equation}
where $D = Eh^3 / 12(1-\nu^2)$ is the plate stiffness and $q(\mathbf{r})$ stands for the vertical distributed force per unit of surface. Consider now an arbitrary  distribution of $N$ linear elastic damped point-like resonators located at positions $\mathbf{R}_\alpha$ with stiffness $K_\alpha$ and damping coefficient $c_\alpha$ and $\alpha=1,\ldots,N$ (see Fig. \ref{fig01}). The force exerted by each resonator at the point $\mathbf{R}_\alpha$ of the plate is
\begin{equation}
  	f(\mathbf{R}_\alpha) = m_\alpha \omega^2 \, \frac{K_\alpha + i \omega \, c_\alpha}{- m_\alpha \omega^2 + i \omega \, c_\alpha + K_\alpha} w\left(\mathbf{R}_\alpha\right) \equiv \hat{K}_\alpha \, w(\mathbf{R}_\alpha).
	\label{eq002}
\end{equation}

The coefficient $\hat{K}_\alpha = \hat{K}_\alpha(\omega)$ is a function of frequency and has units of force per unit of length. It represents the local behaviour of the resonator in the frequency domain.  Other linear models of dissipation may be considered, as those derived from Biot exponential kernel \cite{Biot-1954} or fractional derivatives based viscoelastic models \cite{Bagley-1979}. In the presence of $N$ resonators the general form of the distributed force on the plate is
\begin{equation}
	q(\mathbf{r}) = \sum_{\alpha=1}^{N}   	f(\mathbf{R}_\alpha)  \, \delta(\mathbf{r} - \mathbf{R}_\alpha).
	\label{eq003}	
\end{equation}
Plugging this expression into Eq.~\eqref{eq001} and dividing by the plate stiffness we obtain
\begin{equation}
	\left(\nabla^4 -  k^4 \, \omega^2 \right) w(\mathbf{r}) = \sum_{\alpha=1}^{N}   	t_\alpha  \, w\left(\mathbf{R}_\alpha\right) \, \delta(\mathbf{r} - \mathbf{R}_\alpha),
	\label{eq004}
\end{equation}
where $k = \left(\omega^2 \varrho h / D\right)^{1/4}$ defines the wavenumber in the plate and $t_\alpha = \hat{K}_\alpha(\omega) / D$. To solve Eq.  (\ref{eq004}), first we look for the Green's function of the bare plate by solving
\begin{equation}
	\left(\nabla^4 -  k^4 \, \omega^2 \right) G(\mathbf{r}) = \delta(\mathbf{r}).
	\label{eq005}
\end{equation}
%

The analytical solution is \cite{evans2007penetration}
\begin{equation}
	G(\mathbf{r}) = \frac{i}{8 k^2} \left(\mathcal{H}_0(kr) - \mathcal{H}_0(ikr) \right),
	\label{eq006}
\end{equation}
where $\mathcal{H}_0(kr)$ is the zeroth-order Hankel function of the first kind and $ \mathcal{H}_0(ikr) = -  \frac{2i }{\pi } \mathcal{K}_0(kr)$, with  $\mathcal{K}_0(kr)$ the  zeroth-order modified Bessel function of the second kind. Assuming there is some incident wavefield $\psi_0(\mathbf{r})$,  the total response of the system is solved in terms of the incident and scattered field at the scatterers
\begin{equation}
	w(\mathbf{r}) = \psi_0(\mathbf{r}) + \sum_{\beta=1}^N 	G(\mathbf{r} - \mathbf{R}_\beta) \, t_\beta \, w\left(\mathbf{R}_\beta\right).
\end{equation}
Evaluating this expression at the position of the scatterers, $\mathbf{r} = \mathbf{R}_\alpha, \ \alpha = 1,\ldots,N$, we build the following system of Eqs. \cite{Foldy-1945,Martin-2006,Torrent-2012}
\begin{equation}
	w(\mathbf{R}_\alpha) = \psi_0(\mathbf{R}_\alpha) + \sum_{\beta=1}^N 	G(\mathbf{R}_\alpha - \mathbf{R}_\beta) \, t_\beta \, w\left(\mathbf{R}_\beta\right).
	\label{eq007}
\end{equation}
It can be rearranged in matrix form as
\begin{equation}
	\begin{bmatrix}
		1 - G(\mathbf{0})t_1 			&  -G(\mathbf{R}_1-\mathbf{R}_2)t_2 & \cdots & - G(\mathbf{R}_1-\mathbf{R}_N)t_N   \\
		-G(\mathbf{R}_2-\mathbf{R}_1)t_1 	&  1 - G(\mathbf{\mathbf{0}})t_2 & \cdots & - G(\mathbf{R}_2-\mathbf{R}_N)t_N       \\		
		\vdots								 &  \vdots					& \ddots	&	\vdots						      \\
		-G(\mathbf{R}_N-\mathbf{R}_1)t_1 	&  -G(\mathbf{R}_N-\mathbf{R}_2)t_2 & \cdots & 1 - G(\mathbf{0})t_N 
	\end{bmatrix}
	\begin{Bmatrix}
		w(\mathbf{R}_1) \\
		w(\mathbf{R}_2) \\
		\vdots  \\
		w(\mathbf{R}_N) 
	\end{Bmatrix}
	=
	\begin{Bmatrix}
		\psi_0(\mathbf{R}_1) \\
		\psi_0(\mathbf{R}_2) \\
		\vdots  \\
		\psi_0(\mathbf{R}_N) 
	\end{Bmatrix}.
	\label{eq008}
\end{equation}
By introducing the auxiliary variables
\begin{equation}
	\psi(\mathbf{R}_\alpha)  = \left[1 - G(\mathbf{0}) \, t_\alpha\right] \, w(\mathbf{R}_\alpha)  \quad , \quad  T_\alpha = \frac{ t_\alpha}{1 - G(\mathbf{0}) t_\alpha},	
	\label{eq009}
\end{equation}
%
with $G(\mathbf{0})$ being  the value of the Green's function at the origin,
\begin{equation}
	\lim_{r \to 0} G(kr) = \frac{i}{8 k^2} \equiv G(\mathbf{0}).
	\label{eq012}
\end{equation}
Rewriting Eq. (\ref{eq008}) in terms of the new variables,
\begin{equation}
	\left[\mathbf{I} - \mathbf{g} \right] \, \mathbf{\Psi} = \mathbf{\Psi}_{\text{0}},
	\label{eq010}
\end{equation}
where the matrix $ \mathbf{g} =\mathbf{G} \, \mathbf{T}$ and
\begin{equation}
	\mathbf{G} = 
	\left[
	\begin{array}{cccc}
		0		&			G(\mathbf{R}_1-\mathbf{R}_2)   & \cdots 	& G(\mathbf{R}_1-\mathbf{R}_N) \\
		G(\mathbf{R}_2-\mathbf{R}_1)   &  0			   & \cdots 	& G(\mathbf{R}_2-\mathbf{R}_N)  \\
		\vdots					 & 				\vdots															 &   \ddots  &   \vdots  \\
		G(\mathbf{R}_N-\mathbf{R}_1)			&	 G(\mathbf{R}_N-\mathbf{R}_2)   & \cdots 	& 	0
	\end{array}
	\right] \ , \quad
	\mathbf{T} = 
	\left[
	\begin{array}{cccc}
		T_1		&			\cdots 	&  0 \\
		\vdots					 & 				 \ddots  &   \vdots  \\
		0	&	\cdots 	& 	T_N		 
	\end{array}
	\right].
	\label{eq046}
\end{equation}
The scattering coefficients $\bm{\Psi} = \{\psi(\mathbf{R}_1),\ldots,\psi(\mathbf{R}_N)\}$ emerge from the solution of the above system of equations which in terms of the inverse matrix has the form
\begin{equation}
	\mathbf{\Psi} = \left[\mathbf{I} - \mathbf{g} \right]^{-1} \mathbf{\Psi}_{\text{0}},
	\label{eq047}
\end{equation}
and the total wave field can be obtained from 
\begin{equation}
	w (\mathbf{r})  = \psi_{\text{0}}(\mathbf{r})  + \sum_{\alpha=1}^N G(\mathbf{r}-\mathbf{R}_\alpha) \, T_\alpha \, \psi(\mathbf{R}_\alpha).
	\label{eq048}
\end{equation}
The solution given by Eq.~\eqref{eq047} and the wavefield in Eq.~\eqref{eq048} provides the exact scattering field for the problem at any frequency. \\

\section{Weak scattering approximation}
\label{sec:WeaklyScatteringResonators}

A set of point-like resonators, influenced by both their geometric distribution and mechanical properties relative to the plate, can form a system that, in some cases, results in a  weakly scattered response. For example, in periodic arrangements of point-like scatterers, scattering remains low at frequencies distant from Bragg frequencies, which are determined by the crystal lattice spacing.  Likewise, frequencies far from the  point-like resonators resonance frequencies lead to a reduced scattering response. However, beyond these examples, a rigorous mathematical framework is necessary to systematically classify and differentiate weak scattering. The next section derives conditions for the occurrence of weak scattering, establishing a criterion for selecting lower-order scattering approximations, such as the Born approximation.

\subsection{General conditions for weak scattering and the Born approximation}

As shown in the previous section, the scattering coefficients can be found from the solution of Eq.~\eqref{eq047}. If the elements of $\mathbf{g}$ are sufficiently small, the inverse of the matrix can be efficiently evaluated by means of the Neumann series expansion \cite{Stewart-1998}, namely
\begin{equation}
	\left[\mathbf{I} - \mathbf{g} \right]^{-1} = \mathbf{I} + \mathbf{g} + \mathbf{g}^2 + \cdots = \sum_{n=0}^\infty \mathbf{g}^n.
	\label{eq086}
\end{equation}
%
In this case, the scattering induced by the $N$ resonators is said to be weak. The necessary and sufficient condition to guarantee that such an expansion can be carried out is intimately linked to the magnitude of the spectrum of $\mathbf{g}$ (set of eigenvalues). In fact, if the spectral radius $\rho(\mathbf{g})$ (i.e., the maximum of the absolute values of eigenvalues of $\mathbf{g}$) is less than unity, the convergence of the series is ensured. Moreover, the convergence rate is proportional to the magnitude of $\rho(\mathbf{g})$, provided that $\rho(\mathbf{g})<1$ \cite{Householder-1964,Wilkinson-1988}. Strong scattering is associated to those situations where the spectral radius is $\rho(\mathbf{g})>1$. In those cases, the Neumann series \eqref{eq086} diverges. \\

Assuming that $\rho(\mathbf{g}) < 1$,  the solution can be obtained recursively defining the sequence
\begin{equation}
	\bm{\Psi}_n = \mathbf{g} \,  	\bm{\Psi}_{n-1} \ , \qquad   \bm{\Psi}_0 =  \mathbf{\Psi}_{\text{0}},
	\label{eq087}
\end{equation}
where $\bm{\Psi}_0$ is the incident wavefield. Thus, the scattering coefficients are the result of the series
\begin{equation}
	\mathbf{\Psi} = \left[\mathbf{I} - \mathbf{g} \right]^{-1} \mathbf{\Psi}_{\text{0}} =
	\left(  \mathbf{I} + \mathbf{g} + \mathbf{g}^2 + \cdots \right) \mathbf{\Psi}_{\text{0}} =
	\mathbf{\Psi}_0 + \mathbf{\Psi}_1 + \mathbf{\Psi}_2 + \cdots	= \sum_{n=0}^\infty \bm{\Psi}_n. 
	\label{eq088}
\end{equation}
The above series captures the essence of the multiple scattering behavior of the system, since the effect of  higher order scattered waves are associated to higher order terms of the sequence. The zero-order solution $	\mathbf{\Psi} \approx 	\mathbf{\Psi}_0$ corresponds to 
\begin{equation}
	\psi(\mathbf{R}_\alpha) \approx \psi_0(\mathbf{R}_\alpha).
	\label{eq0106}
\end{equation}
The total wavefield from Eq.~\eqref{eq048} can be simply written as
\begin{equation}
	w (\mathbf{r})  \approx \psi_{\text{0}}(\mathbf{r})  + \sum_{\alpha=1}^N G(\mathbf{r}-\mathbf{R}_\alpha) \, T_\alpha \, \psi_0(\mathbf{R}_\alpha).
	\label{eq011}
\end{equation}
Higher order scattering waves are neglected in this expression. As an extension of the concept used in quantum mechanics, the so-obtained wavefield  is called {\em Born--approximation} in the context of  2D flexural waves in thin elastic plates. As shown in  Eq.  \eqref{eq011}, the scattered wavefield of Born-approximation is proportional to the  strength) of the scatterers and also depends on their relative positions via the Green matrix. Adding terms of the series in Eq.~\eqref{eq088} to the computation of the scattering field is equivalent to consider higher order scattering waves. The accuracy of the Born-approximation is closely related to the magnitude of the spectral radius $\rho(\mathbf{g})$, which is a closed-form dimensionless measurement of the scattering intensity. Thus, lower values of  $\rho(\mathbf{g})$ will be associated with good agreement of the wavefield approximated by Eq.~\eqref{eq011} as will be validated later in the numerical examples. In the following we will propose an approximation of the spectral radius $\rho(\mathbf{g})$, given that it corresponds to the maximum eigenvalue of the matrix in magnitude. This approximation will allow us to dissociate between two relevant properties of scattered media: (i) the relative position between scatterers and (ii) the pointwise scattering intensity of each resonator. \\

%

Some asymptotic expressions based on the far-field approximations of the Hankel function can be derived, which will allow us to relate the wave dispersion to the geometric and physical properties of the scatterers' distribution. As known, the term $\mathcal{H}_0(kr)$ in the Green's function represents the oscillatory wavefield, while the function $\mathcal{K}_0(kr)$ encapsulates the envanescent wavefield. It is expected that after some distance from the source, the latter vanishes and thus can be neglected. The far field approximation of the zero-order Hankel function is
\begin{equation}
		\mathcal{H}_0(kr) \approx \frac{1-i}{\sqrt{\pi k r}} e^{ikr}  \ , \qquad
		\mathcal{H}_0(ikr) \approx - \sqrt{\frac{2}{\pi k r}} e^{-kr}  		
		\quad , \qquad kr > 2 ,
		\label{eq013}
\end{equation}
Therefore, we can approximate 
\begin{equation}
		G(\mathbf{r}) = 
		\frac{i}{8 k^2} \left(\mathcal{H}_0(kr) - \mathcal{H}_0(ikr) \right) \approx 
		\frac{i}{8 k^2} \left( \frac{1-i}{\sqrt{\pi k r}} e^{ikr}  + \sqrt{\frac{2}{\pi k r}} e^{-kr}  		 \right) 	\approx \frac{G_p}{\sqrt{kr}} e^{ikr} 
		\quad , \qquad kr > 3 ,
		\label{eq014}
\end{equation}
where $G_p = \frac{i (1-i)}{8 k^2 \sqrt{\pi} }$.   \\

In Eq.  \eqref{eq048}, the second term represents the scattered wave field, which will be denoted by $w_s(\mathbf{r})$. By considering the Born approximation, the scattered field can be approximated by
\begin{equation}
	w_s(\mathbf{r}) \approx  \sum_{\alpha=1}^N G(\mathbf{r}-\mathbf{R}_\alpha) \, T_\alpha \, \psi_0(\mathbf{R}_\alpha).
		\label{eq015}
\end{equation}
%




Let us consider a plane wave as incident field, with direction given by vector $\mathbf{k}_0$, i.e.
\begin{equation}
    \psi_0(\mathbf{r}) = e^{i\mathbf{k}_0 \cdot \mathbf{r}}.
    \label{eq024}
\end{equation}
In this case, the far-field approximation of the Green function together with the Born-approximation leads to a closed form expression of the scattered field. Thus, assuming that $k\left|\mathbf{r} - \mathbf{R}_\alpha \right| \gg 1$, we can apply Eq.~\eqref{eq014}, resulting
\begin{equation}
	G(\mathbf{r} - \mathbf{R}_\alpha) \approx 
	G_p \, \frac{e^{ik\left|\mathbf{r} - \mathbf{R}_\alpha \right| } }{\sqrt{k\left|\mathbf{r} - \mathbf{R}_\alpha \right|}} 
	\approx G_p \, \frac{e^{ikr} }{\sqrt{kr}}  e^{- ik \mathbf{u}_{\mathbf{r}} \cdot \mathbf{R}_\alpha},
	\label{eq017}
\end{equation}
where $\mathbf{u}_{\mathbf{r}} = \left(\cos \theta, \sin \theta \right)$ denotes the unitary vector in the direction of $\mathbf{r}$. Plugging Eqs. \eqref{eq017} and \eqref{eq024} into Eq.~\eqref{eq015} , we find
\begin{equation}
	w_s(\mathbf{r}) \approx 
	\sum_{\alpha=1}^N G_p \, \frac{e^{ikr} }{\sqrt{kr}}  e^{- ik \mathbf{u}_{\mathbf{r}} \cdot \mathbf{R}_\alpha} \, T_\alpha \, e^{\mathbf{k}_0 \cdot \mathbf{R}_\alpha} = 
	\frac{e^{ikr} }{\sqrt{kr}}  \sum_{\alpha=1}^N \left(G_p T_\alpha \right)  \,  e^{- i \left(k \mathbf{u}_{\mathbf{r}} - \mathbf{k}_0\right) \cdot \mathbf{R}_\alpha}  	,
	\label{eq018}
\end{equation}
where $e^{- ik \mathbf{u}_{\mathbf{r}} \cdot \mathbf{R}_\alpha}$ represents the scattered wave with direction $\theta$ at the position of the scatterer $\mathbf{R}_\alpha$. Thus, the scattered wavevector in the direction $\theta$ will be denoted by $\mathbf{k}_s= k \, \mathbf{u}_{\mathbf{r}} = k  \left(\cos \theta, \sin \theta \right)$. Both $\mathbf{k}_s$ and $\mathbf{k}_0$ are propagating wavevectors for a particular frequency $\omega$, therefore both share the same magnitude although different directions, verifying that
\begin{equation}
	\left|\mathbf{k}_s \right| = 	\left|\mathbf{k}_0 \right| = k =  \left(\frac{\omega^2 \varrho h}{D} \right)^{1/4}.
	\label{eq020}
\end{equation} 
\begin{figure}[h]%
	\begin{center}
		\begin{tabular}{c}
			\includegraphics[width=7cm]{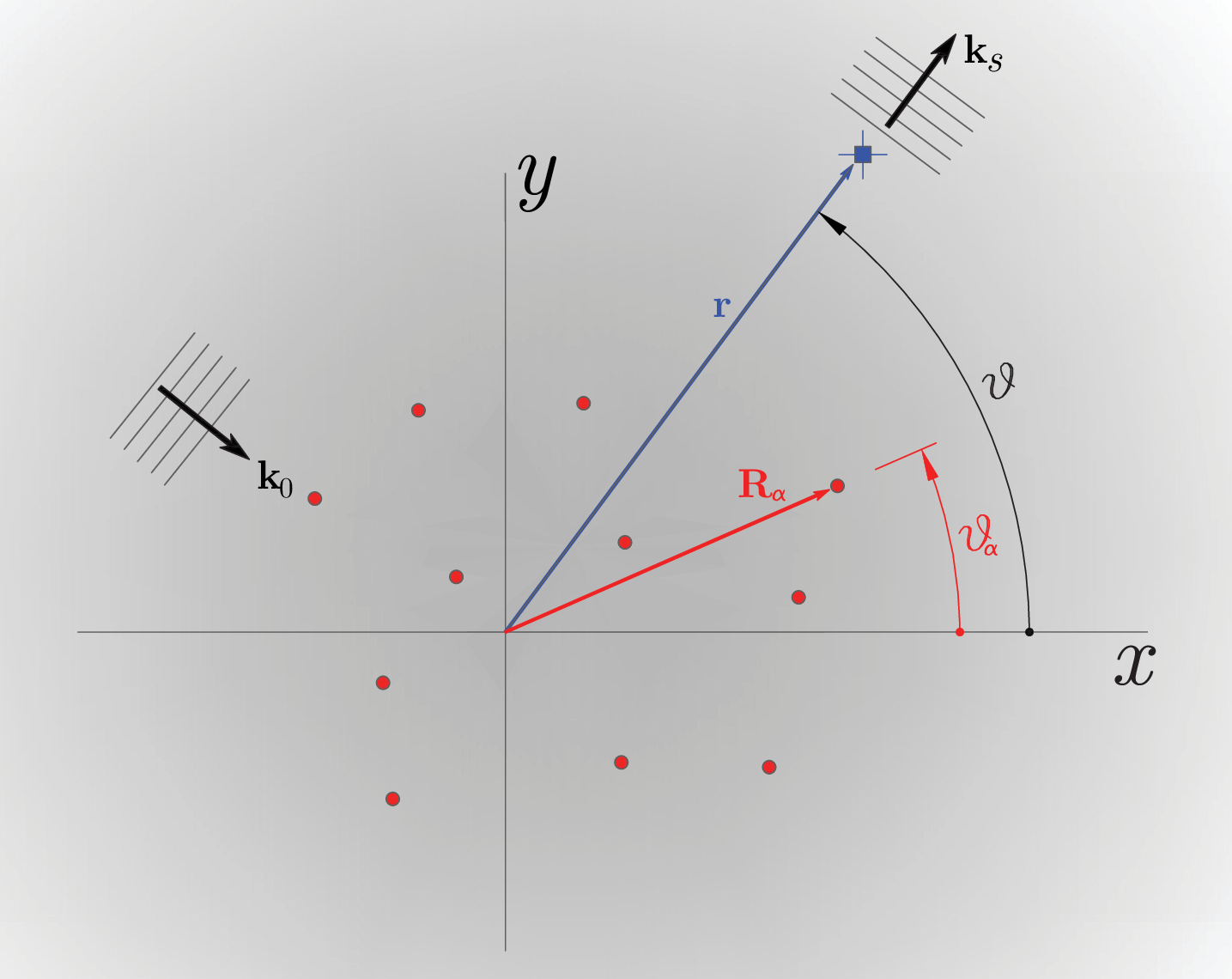} 
		\end{tabular}			
		\caption{Plate with a distribucion of $N$ point-like resonators at coordinates $\mathbf{R}_\alpha$. Incident wave with wavenumber $\mathbf{k}_0$. The vector $\mathbf{k}_s$ represents the scattered field on the direction $\theta$}%
		\label{fig03}%
	\end{center}
\end{figure}

Figure \ref{fig03} schematically shows all these directions together with the incident field. The non-dimensional parameter
 \begin{equation}
 	\tau_\alpha = G_p T_\alpha = \frac{G_p t_\alpha }{1 - G(\mathbf{0}) t_\alpha} = \frac{1-i}{\sqrt{\pi}} \, \frac{G(\mathbf{0}) t_\alpha }{1 - G(\mathbf{0}) t_\alpha},
 	\label{eq025}
 \end{equation}
 is a local measurement of the scattering produced by a single resonator. Eq.  \eqref{eq018} can then be explicitly expressed in terms of these parameters, on the positions of scatterers $\mathbf{R}_\alpha$ and on the difference between the incident and scattered wave, yielding
\begin{equation}
			w_s(\mathbf{r}) \approx 
		\frac{e^{ikr} }{\sqrt{kr}} \sum_{\alpha=1}^N \tau_\alpha  \,   e^{- i \left( \mathbf{k}_s - \mathbf{k}_0 \right) \cdot \mathbf{R}_\alpha}  \equiv \frac{e^{ikr} }{\sqrt{kr}} {F} \left( \mathbf{k}_s - \mathbf{k}_0 \right),
		\label{eq019}
\end{equation}
where the function $F(\mathbf{K})$ is defined as
\begin{equation}
		F(\mathbf{K}) =  \sum_{\alpha=1}^N \tau_\alpha  \,   e^{- i  \mathbf{K} \cdot \mathbf{R}_\alpha}  .
 	\label{eq026}		
\end{equation}
If all resonators share the same dynamical properties $\tau_\alpha \equiv \tau_0$, for $0 \leq \alpha \leq N$, the function $F(\mathbf{K})$ can then be written as
\begin{equation}
		F(\mathbf{K}) =  \tau_0 \sum_{\alpha=1}^N   \,   e^{- i  \mathbf{K} \cdot \mathbf{R}_\alpha}  = \tau_0 \, \mathcal{FT}   \left\{\sum_\alpha \delta(\mathbf{r-\mathbf{R}_\alpha }) \right\},
\end{equation}


%
where $ \mathcal{FT}   \left\{ \bullet \right\} $ denotes the Fourier transform of a space-defined function. Thus, the right-hand term represents the Fourier coefficients in the reciprocal space $\mathbf{K}$ corresponding to the spatial direct domain of the medium defined by $N$ particles at positions $ \{\mathbf{R}_\alpha\}$, which mathematically can be defined as a sum of Dirac-delta functions (discrete points in space), i.e. $d(\mathbf{r}) = \sum_\alpha \delta(\mathbf{r-\mathbf{R}_\alpha })$. According to Eq.  \eqref{eq019}, the scattered wave is proportional to the Fourier coefficients, evaluated on the vectors $\mathbf{K} = \mathbf{k}_s-\mathbf{k}_0$. Here is a direct relationship between the information of the reciprocal space of the set of points and the physics of the problem. It is known that in periodic media, the von Laue condition \cite{Ashcroft-1976,Kittel-2005} implies that constructive interference will occur if the difference between the incident and reflected wave is a vector of the reciprocal space, say  $\mathbf{K} = \mathbf{k}_s-\mathbf{k}_0$.  Furthermore, in the general case of resonators with different properties, expression \eqref{eq026} allows us to interpret the parameter $\tau_\alpha$ as an atomic factor associated with the $\alpha$-th resonator. Thus, Eq.  \eqref{eq026} extends the known concept of Structure Factor \cite{Ashcroft-1976,Kittel-2005,RomeroGarcia-2019} to wave mechanics in plates. The dispersion relation of the plate together with the interpretation given by the Oswald sphere \cite{RomeroGarcia-2021}, allows us to assign a frequency to each vector of the reciprocal lattice. It is therefore possible to define the structure factor of the plate as function of the reciprocal space lattice as
\begin{equation}
	S(\mathbf{K}) = \frac{1}{N} \left| 		F(\mathbf{K}) \right|^2.
	\label{eq027}
\end{equation}
This direct relationship between physical space, resonator properties and scattering intensity could not be established without the Born approximation. In turn, we have proved above the direct relationship between the latter and the spectral radius of the scattering matrix, $\rho(\mathbf{g})$.  The next section explores how $\rho(\mathbf{g})$ is influenced separately (in a multiplicative manner) by both the local properties of the resonators and their global spatial distribution, linking real physical properties with the mathematical condition of weak scattering. 


\section{Physical interpretation of the spectral radius}

The mathematical conditions governing weak scattering in plates and their relation to the scattering matrix have been derived in the previous section. From the physical point of view, scattering is due to two fundamental factors of the collection of scatterers: (i) their individual frequency response functions, meaning their relative inertial characteristics and resonant frequency respect to the properties of the plate, and (ii) the correlation between the distances of the scatterers and their relation to the wavelength. These two properties are typically exploited for the design of metamaterials by aligning Bragg frequencies with resonant frequencies. By invoking the far-field approximations derived above, we observe that the information from the two aforementioned factors is encapsulated within the scattering matrix. Moreover, the spectral radius can be approximated in such a way that both factors are separated, appearing as the product of two independent spectral radii. \\

Indeed, considering that resonators are well separated each other, we can assume Eq.  \eqref{eq017}. Then,  the terms in the scattering matrix $\mathbf{g}$ can be written as
\begin{equation}
	g_{\alpha \beta} = G(\mathbf{R}_\alpha - \mathbf{R}_\beta) T_\beta 
	\approx G_p \, T_\beta \, \frac{e^{ik\left|\mathbf{R}_\alpha  - \mathbf{R}_\beta \right| } }{\sqrt{k\left|\mathbf{R}_\alpha - \mathbf{R}_\beta \right|}}
	\equiv \tau_\beta \, \mathcal{G}_{\alpha \beta} \quad \text{for} \  1 \leq \alpha, \beta \leq N ,
	\label{eq021}
\end{equation}


%
The dimensionless magnitude  $\mathcal{G}_{\alpha \beta}$ depends exclusively on the relative distance between scatterers with respect to the wavelength, denoted as $k \, r_{\alpha \beta} = k \left|\mathbf{R}_\alpha - \mathbf{R}_\beta \right|$. Thus, we can define the matrices $\bm{\mathcal{G}}$ and $\bm{\tau}$ as
\begin{equation}
\bm{\mathcal{G}}= 
	\left[
	\begin{array}{cccc}
		0		&			e^{ik r_{12}}/\sqrt{kr_{12}}    & \cdots 	&  e^{ik r_{1N}}/\sqrt{kr_{1N}} \\
		e^{ik r_{12}}/\sqrt{kr_{12}}   &  0			   & \cdots 	& e^{ik r_{2N}}/\sqrt{kr_{2N}}  \\
		\vdots					 & 				\vdots															 &   \ddots  &   \vdots  \\
		e^{ik r_{1N}}/\sqrt{kr_{1N}}  	&	e^{ik r_{2N}}/\sqrt{kr_{2N}}     & \cdots 	& 	0
	\end{array}
	\right] \ , \quad
	\bm{\tau} = 
	\left[
	\begin{array}{cccc}
		\tau_1		&			\cdots 	&  0 \\
		\vdots					 & 				 \ddots  &   \vdots  \\
		0	&	\cdots 	& 	\tau_N		 
	\end{array}
	\right].
	\label{eq022}
\end{equation}
The scattering matrix can be approximated for weak scattering in the far-field by $\mathbf{g} \approx \bm{\mathcal{G}} \, 	\bm{\tau}$. In general, the spectral radius of the product of matrices cannot be bounded by the product of their respective spectral radii. However, in practice, this product works quite well as an upper-bound approximation of the exact spectral radius in this specific case, where one of the matrices is diagonal. Therefore, we can approximately express this as follows: 

\begin{equation}
	\rho(\mathbf{g} ) \approx \rho(\bm{\mathcal{G}} ) \, \rho(\bm{\tau})  =  \rho(\bm{\mathcal{G}} ) \, \max_{\alpha} \left| \tau_\alpha \right| .
		\label{eq023}
\end{equation}

In practice, the right-hand side is usually an upper bound of the exact spectral radius. In the case of having a distribution of scatterers with equal mechanical properties, the above expression is not an approximation of the spectral radius, but an exact expression. The information about the position of the scatterers relative to the wave with wavenumber \(k\) is captured for each frequency in the spectral radius \(\rho(\bm{\mathcal{G}})\), while the relative impedance between each scatterer and the plate is frequency-dependent in the expression \(\rho(\bm{\tau})\). Both factors are multiplicative, which helps to explain the presence of strong or weak scattering for each frequency. In the numerical examples presented in the following section, a comparison between these two factors will be analysed for a given collection of scatterers and for a frequency range in the host medium.
%
%
%
%

\section{Numerical example 1: periodic arrangement of scatterers}
\label{sec:NumericalExamples}

\begin{figure}[!]%
	\begin{center}
		\begin{tabular}{c}
			\includegraphics[width=6cm]{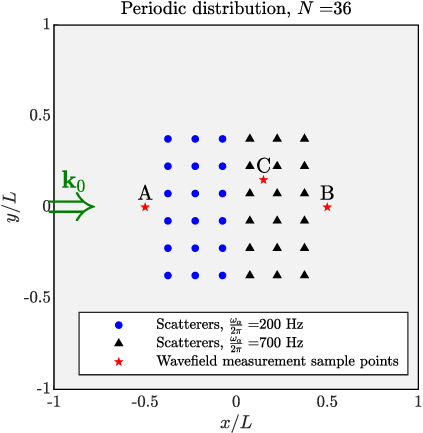} 
		\end{tabular}			
		\caption{Distribution of scatterers in Example 1. Two families of scatterers with two different resonances: blue-circle and black-triangle scatterers. Incident wave rightwards with wavenumber $\mathbf{k}_0$. Three measurement points: \textbf{A}, \textbf{B} and \textbf{C}. The wavefield response is shown in Fig.  \ref{fig05}}%
		\label{fig04}%
	\end{center}
\end{figure}

As first application, we will study the wave response of an aluminum thin plate with a periodic arrangement of scatterers. The plate has density $\varrho = 2.7$ t/m$^3$, Young modulus $E=70$ GPa, Poission coefficient $\nu = 0.3$ and thickness $h=2$ mm.  We consider $N=36$ scatterers separated a regular distance of $s=30$ cm. The scatterers have a mass of $m_\alpha = 5$ g and the set is divided into two groups as shown in Fig. \ref{fig04}, with natural frequencies $\frac{\omega_\alpha }{2\pi} =200$ Hz and $\frac{\omega_\alpha }{2\pi} =700$ Hz, respectively. \\

\begin{figure}[!h]%
	\begin{center}
		\begin{tabular}{ccc}
			\includegraphics[width=8cm]{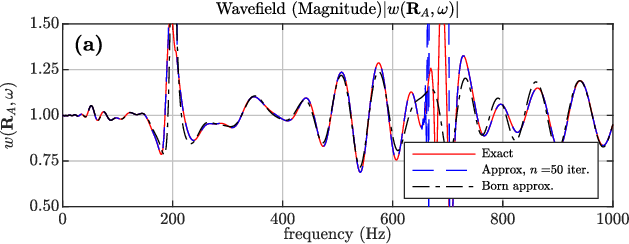} &
			\includegraphics[width=8cm]{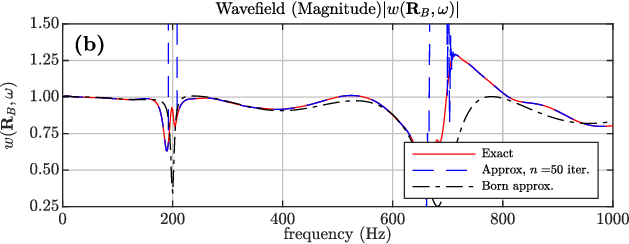}  \\
			\includegraphics[width=8cm]{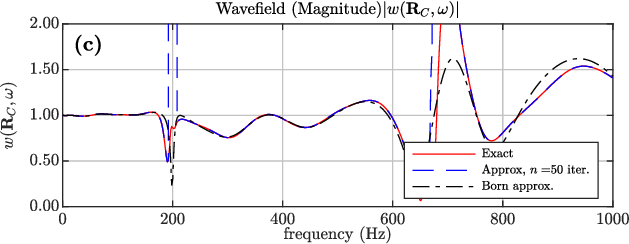} &
			\includegraphics[width=8cm]{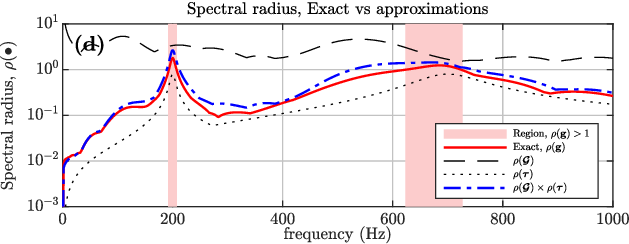} 			
		\end{tabular}			
		\caption{(a, b, c) Wavefields at sample points \textbf{A}, \textbf{B}, \textbf{C} shown in Fig. \ref{fig04}. (d) Spectral radius of scattering matrix vs. frequency. Shaded in the is shown the region of frequency where the spectral radius verifies $\rho(\mathbf{g})<1$, regions which coincide with the natural frequencies used for the point-like resonators}%
		\label{fig05}%
	\end{center}
\end{figure}

Given the position of the resonators on the plate and their mechanical properties, the scattering matrix \(\mathbf{g}\) can be determined for a range of frequencies. It is therefore possible to know the spectral radius of this matrix in advance and distinguish in the spectrum those regions that we can consider to have weak scattering (\(\rho(\mathbf{g}) < 1\)) from those with strong scattering (\(\rho(\mathbf{g}) > 1\)). In Fig. \ref{fig05} (top-left), the spectral radius \(\rho(\mathbf{g})\) is plotted as a function of frequency, and the regions of high scattering are highlighted. It is important to note that this measure is independent of the type of incident wave as well as its direction, and depends solely on the properties of the scatterer arrangement relative to those of the plate.  The spectral radii of the matrices \(\bm{\mathcal{G}}\) and \(\bm{\tau}\), defined in Eqs. (22), are also {shown}. The distribution of \(\rho(\bm{\mathcal{G}})\) only depends on the relative distances between resonators with respect to the wavelength. Some clear maximum values are observed around the  frequencies associated with the lattice spacing (Bragg frequencies). The maximum values of \(\rho(\bm{\tau})\) are located at the resonance frequencies of the scatterers. Altogether, the product of both spectral radii is not the exact value but comes quite close, thereby allowing the scattering phenomenon to be explained in a separate and dimensionless  manner. \\

Figures \ref{fig05} (a-c) show the wave field amplitude at three points on the plate with an incident wave $\psi(\mathbf{r} = e^{k_0 \cdot x})$. Both the chosen points, labeled A, B, and C, and the wave direction $\mathbf{k}_0 = \{k_0,0\}$ are shown in Fig. \ref{fig04}. To determine the wave field, the exact method based on the inversion of the matrix \(\mathbf{I-g}\) was used, as well as the method based on Neumann series with \(n=50\) iterations. This number of iterations is more than  enough to observe convergence in regions with a spectral radius less than one, see Fig. \ref{fig05} (d) . However, it clearly shows divergence in regions of strong scattering (shaded red areas in the spectral radius graph). The Born approximation, based one single iteration of the Neumann series, Eq.  \eqref{eq011}, has also been plotted along the frequency range, showing very good agreement in the regions of weak scattering. \\

\begin{figure}[h]%
	\begin{center}
		\begin{tabular}{c}
			\includegraphics[width=11cm]{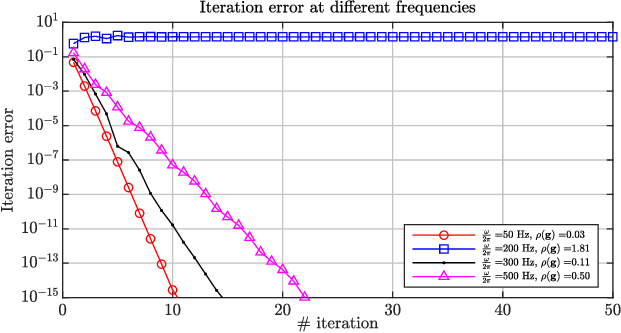} 
		\end{tabular}			
		\caption{Error in the prediction of the wave field for three different frequencies using the Neumann series expansion of the response. It is expected that when the spectral radius is less than one, the difference between the response in one iteration and the next approaches zero with linear convergence, as shown for the frequencies of 50, 300, and 500 Hz, with a speed inversely proportional to the value of \(\rho(\mathbf{g})\). The frequency associated to  a spectral radius greater than one, i.e. 200 Hz,  does not converge.}%
		\label{fig08}%
	\end{center}
\end{figure}

The convergence of the Neumann series is straightforward and is based on obtaining each solution from the previous one by multiplying the scattering matrix as shown in Eq. \eqref{eq087}. Therefore, it is expected that the error in the iteration will be linear when the spectral radius is less than one. Additionally, the slope of the error will be proportional to this spectral radius, as shown in Fig. \ref{fig08}, where the error pattern after several iterations is represented. It can be seen that in the case of weak scattering, only a few iterations are actually necessary to achieve a good estimate. \\

\begin{figure}[!h]%
	\begin{center}
		\begin{tabular}{cccc}
			\includegraphics[width=6.0cm]{figures/Example01_sketch.eps} & 
			\includegraphics[width=6.0cm]{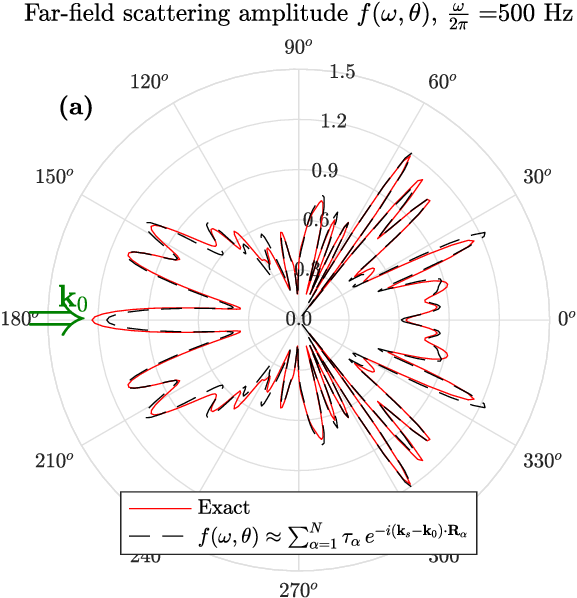}	\\ \\
			\includegraphics[width=7.0cm]{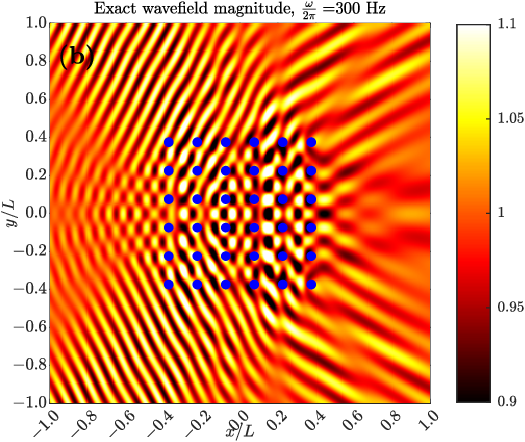}  &
			\includegraphics[width=7.0cm]{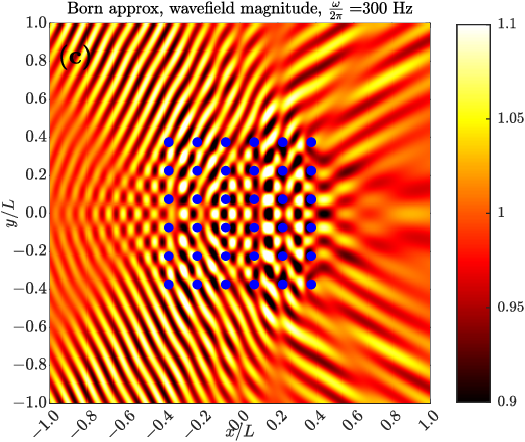}  
		\end{tabular}			
		\caption{Scattering from a periodic array of $N=36$ point-like resonators (blue-circles with natural frequency $\frac{\omega_\alpha}{2\pi}=200$ Hz and black-triangles with $\frac{\omega_\alpha}{2\pi}=700$ Hz. (a) Far-field scattering amplitude at \(\frac{\omega}{2\pi} =500\) Hz, comparing the exact solution with the Born approximation. (b) Exact wavefield magnitude at \(\frac{\omega}{2\pi} =300\) Hz. (c) Wavefield magnitude using the Born approximation at \(\frac{\omega}{2\pi} =300\) Hz.}%
		\label{fig06}%
	\end{center}
\end{figure}

According to our theoretical developments, the evaluation of the response of a plate at a frequency associated with weak scattering can be approximated using the Born approximation, and the result is expected to be similar to the exact one in all regions of the plate and in all directions. To validate this statement, let us examine how the entire plate responds to an incident wave using the Born approximation, and we will calculate the scattering amplitude in the far field and the wave field magnitude at all points on the plate. Figure \ref{fig06} (top-right) shows the magnitude of the function \(f(\omega,\theta)\) for a frequency of 500 Hz, along with the exact solution obtained with multiple scattering. The result shows that, as predicted by theory, there is a strong similarity for all scattering directions. Additionally,  Fig. \ref{fig06} (bottom) shows the magnitude of the wave field at a frequency of 300 Hz, evaluated both exactly (multiple scattering) and using the Born approximation. Excellent agreement is observed between both solutions, not only in regions outside the scatterers cluster but also in the inner regions, as the spectral radius criterion is global for the entire plate and depends on the frequency.

\section{Numerical example 2: random arrangement of scatterers}

\begin{figure}[ht]%
	\begin{center}
		\begin{tabular}{cccc}
			\includegraphics[height=5.0cm]{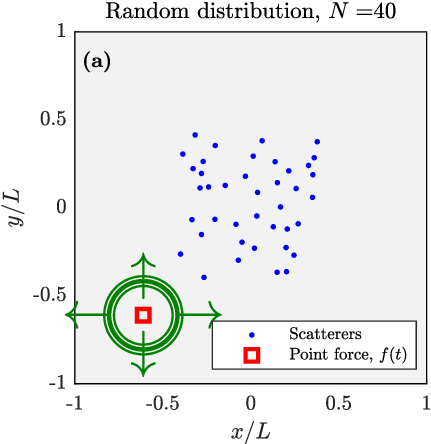} &
			\includegraphics[height=5.0cm]{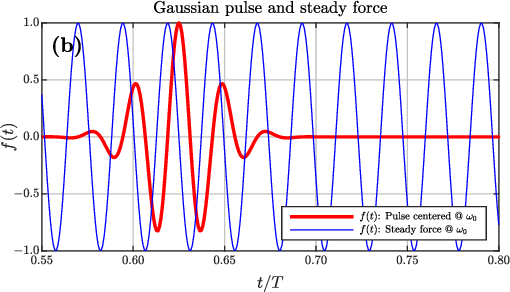}  
		\end{tabular}			
		\caption{(a) Random distribution of scatterers with resonance frequency $\frac{\omega_\alpha }{2\pi} =1300$ Hz and mass $m_\alpha = 10$ g. (b) Two input point-forces: steady of frequency $\frac{\omega_0}{2\pi}=850$ Hz and Gaussian pulse with spectrum centered at $\omega_0$}%
		\label{fig09}%
	\end{center}
\end{figure}
In this new example, we will simulate a set of \(N=40\) resonators randomly arranged on a plate. The mass of the resonators is \(m_\alpha = 10\) g, and the resonance is \textcolor{black}{\(\frac{\omega_\alpha }{2\pi} =1300\) Hz}. In this case, we will consider an incident wave resulting from the action of a point force on the plate, the location of which is shown in Fig.~\ref{fig09}(a), along with the positions of the scatterers. We will perform two simulations using two types of incident field: one stationary at the frequency \textcolor{black}{\(\frac{\omega_0}{2\pi} =850\) Hz}, and the other a pulse with a Gaussian distribution in the frequency spectrum, centered at the frequency \(\omega_0\) and with a standard deviation of \(0.2 \omega_0\). The temporal shape of both inputs is shown in Fig. \ref{fig09} (b).\\
\begin{figure}[h]%
	\begin{center}
		\begin{tabular}{c}
			\includegraphics[width=12cm]{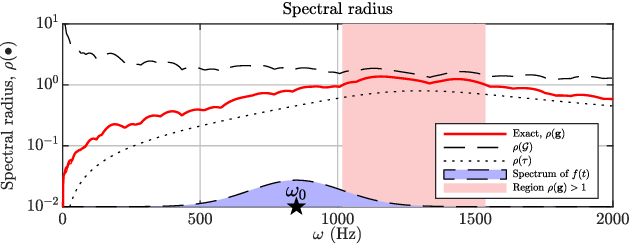} 
		\end{tabular}			
		\caption{Spectral radius of arrangement of scatterers along the frequency range. In blue, the region corresponding to the frequency spectrum of the Gaussian pulse-force is represented, centered at $\frac{\omega_0}{2\pi}=850$ Hz.}%
		\label{fig10}%
	\end{center}
\end{figure}

The evolution of the spectral radius is represented in Fig.~\ref{fig10}, along with the value associated with the impedance of each individual scatterer, \(\rho(\bm{\tau})\), and the effect of the distribution, given by \(\rho(\bm{\mathcal{G}})\). The combination of these properties leads to a region of strong scattering between 1000 Hz and 1500 Hz, mainly due to the effect of resonance. Alongside these curves, the region where the frequencies of the wave pulse spectrum from the point force are located is shown, centered at \textcolor{black}{\(\frac{\omega_0}{2\pi} =850\) Hz}. \\

\begin{figure}[h]%
	\begin{center}
		\begin{tabular}{cccc}
			\includegraphics[width=7.0cm]{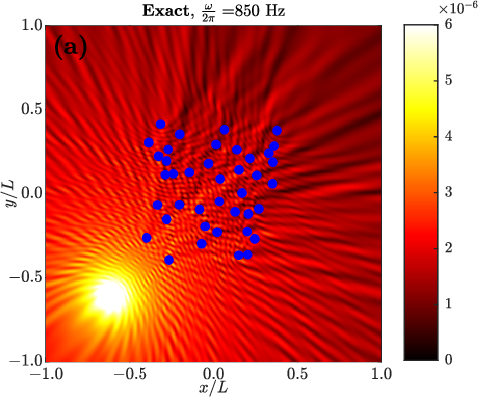}  &
			\includegraphics[width=7.0cm]{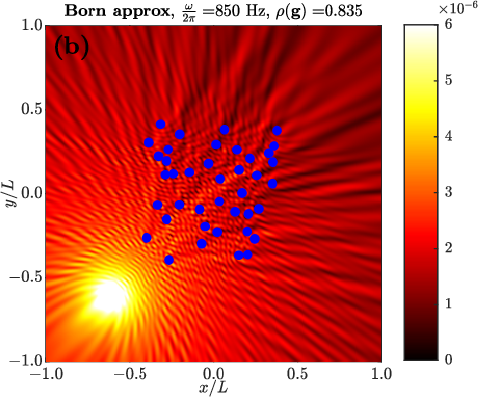}  
		\end{tabular}			
		\caption{Wavefield magnitude at the plate for steady point force at source with frequency $\frac{\omega_0}{2\pi} =850$ Hz. (a) Exact solution based on multiple scattering theory. (b) Born approximation}%
		\label{fig11}%
	\end{center}
\end{figure}

As verified in the previous example, the plate's response evaluated with the Born approximation for a plane wave generally yields good scattering field results when the spectral radius is less than one. In this example, the response to an incident stationary point-source wave has been simulated, and the results are shown in Fig. \ref{fig11} for the excitation frequency \textcolor{black}{\(\frac{\omega_0}{2\pi} =850\) Hz}, for which the spectral radius is \(\rho(\mathbf{g}) = 0.835\). The results show excellent agreement with the exact solution based on multiple scattering.\\

\begin{figure}[h]%
	\begin{center}
		\begin{tabular}{cccccc}
			 																																			&
			\includegraphics[width=2.7cm]{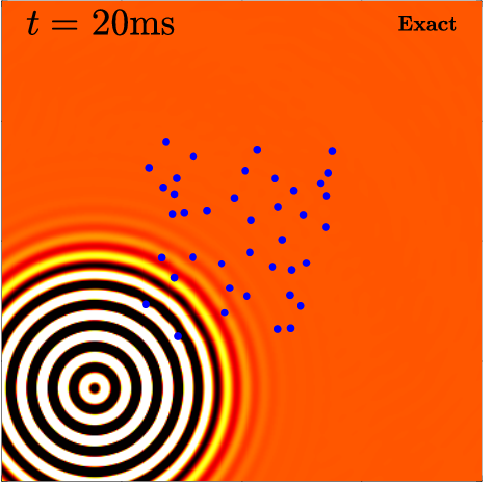}  &
			\includegraphics[width=2.7cm]{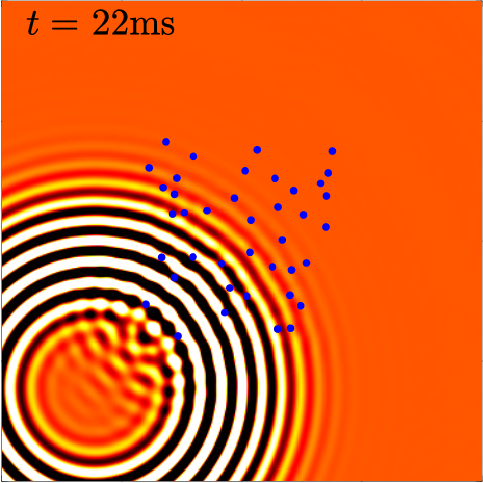}   &
			\includegraphics[width=2.7cm]{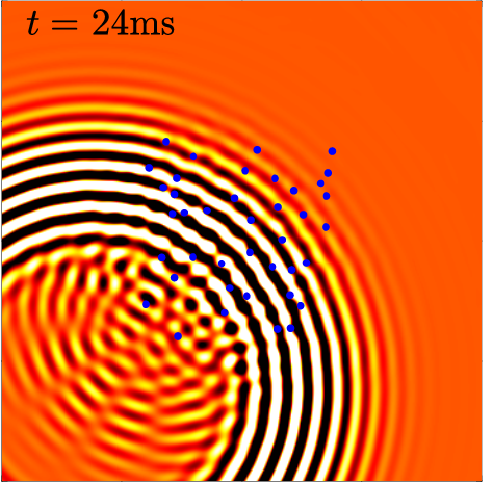}  &
			\includegraphics[width=2.7cm]{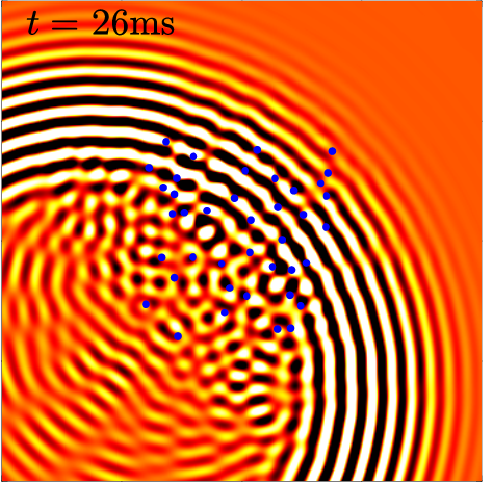}  &
			\includegraphics[width=2.7cm]{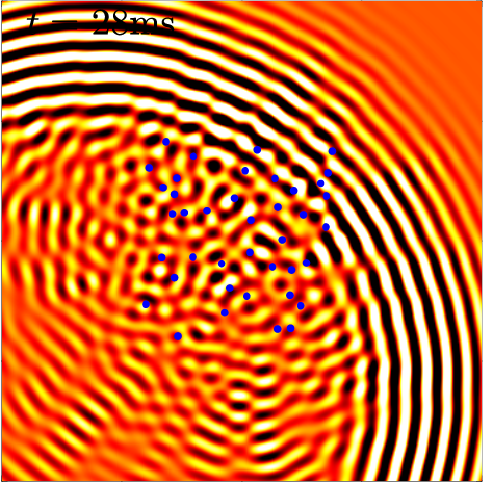}  \\ \\																			
			\rotatebox{90}{\textbf{Exact (MST)}}																					&
			\includegraphics[width=2.7cm]{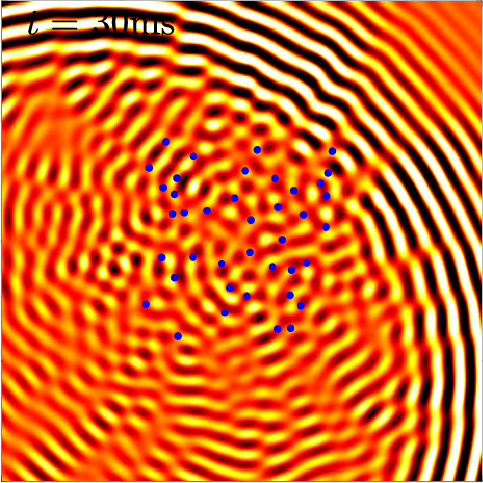}  &
			\includegraphics[width=2.7cm]{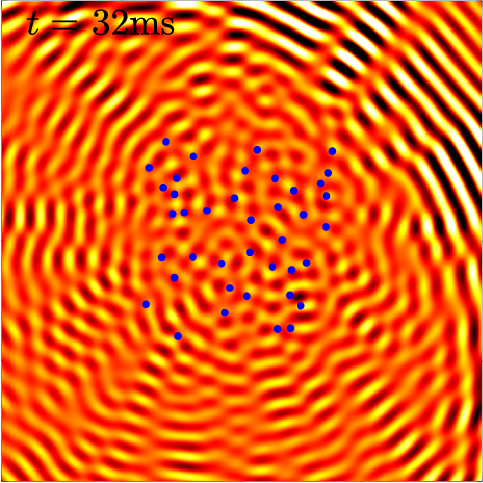}   &
			\includegraphics[width=2.7cm]{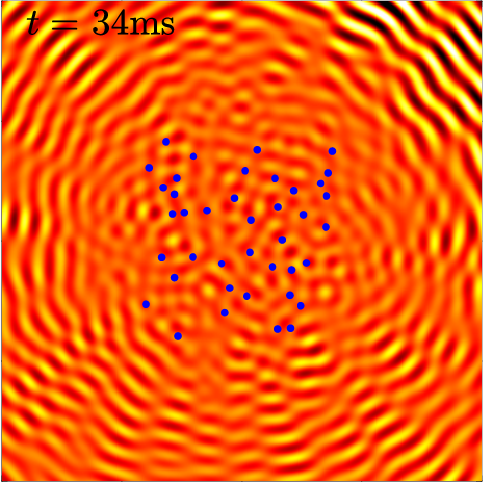}  &
			\includegraphics[width=2.7cm]{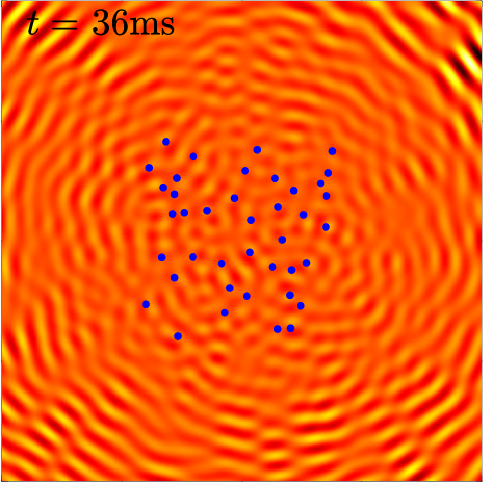}  &
			\includegraphics[width=2.7cm]{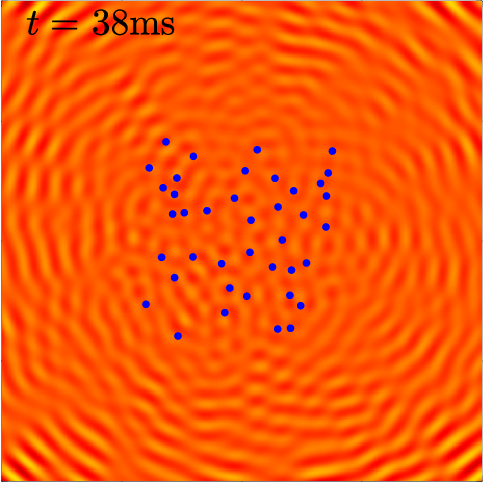}  \\  \\ 
																																						 &
			\includegraphics[width=2.7cm]{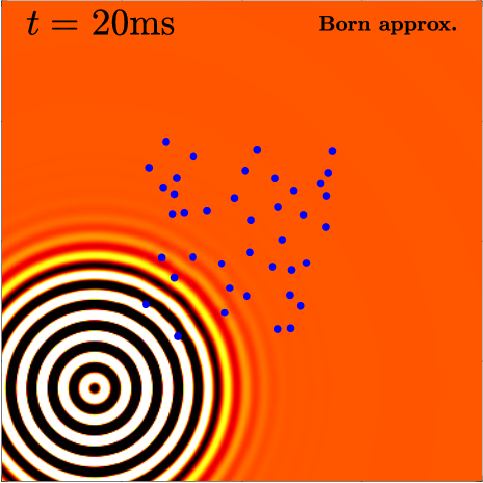}  &
			\includegraphics[width=2.7cm]{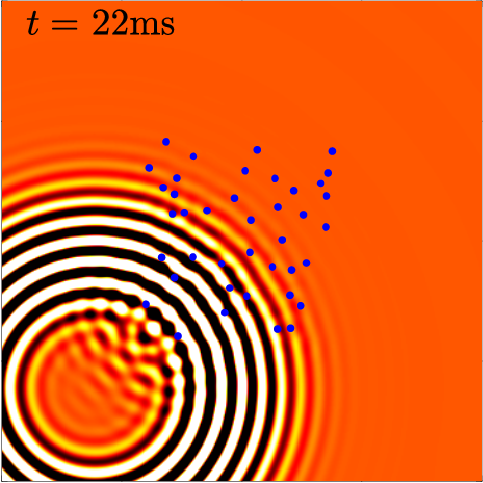}   &
			\includegraphics[width=2.7cm]{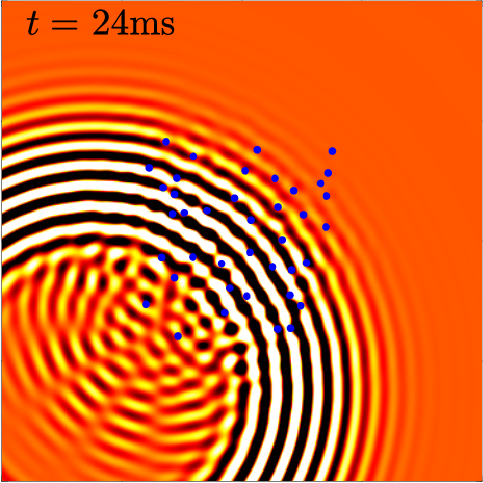}   &
			\includegraphics[width=2.7cm]{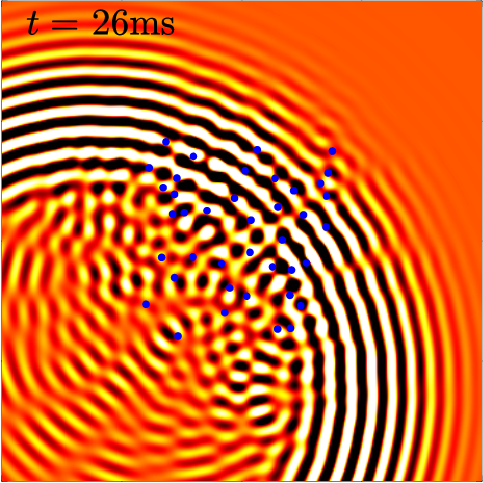}   &
			\includegraphics[width=2.7cm]{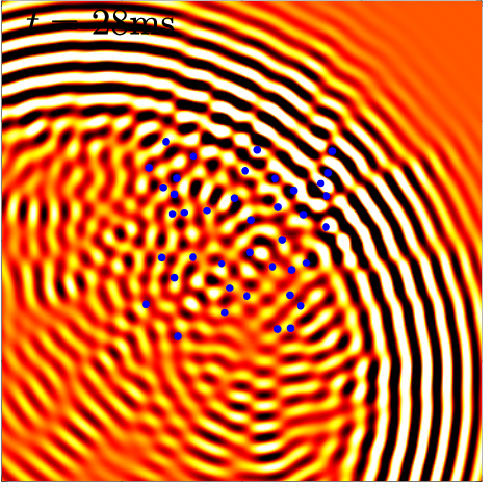}   \\ \\
			\rotatebox{90}{\textbf{Born approx.}} 																				&
			\includegraphics[width=2.7cm]{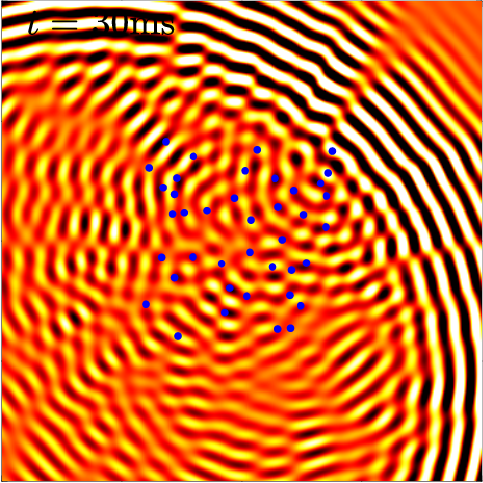}  &
			\includegraphics[width=2.7cm]{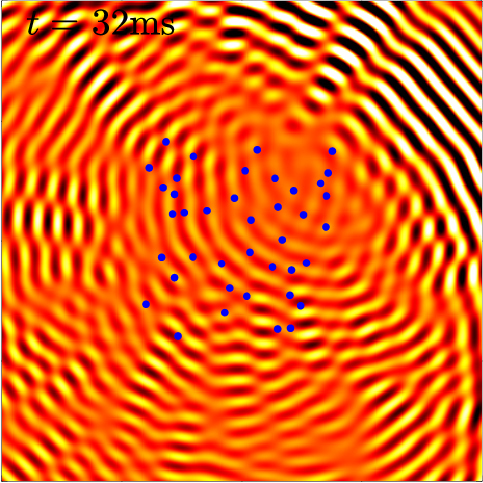}   &
			\includegraphics[width=2.7cm]{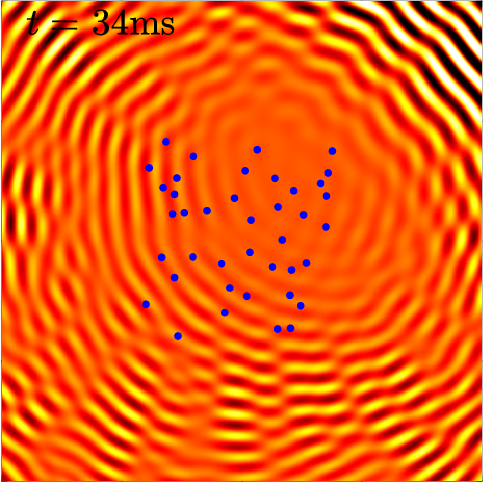}   &
			\includegraphics[width=2.7cm]{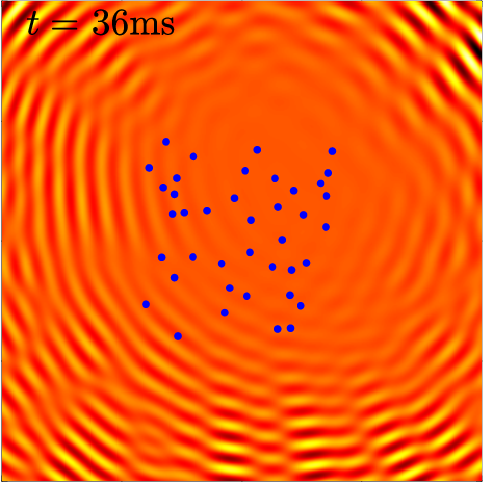}   &
			\includegraphics[width=2.7cm]{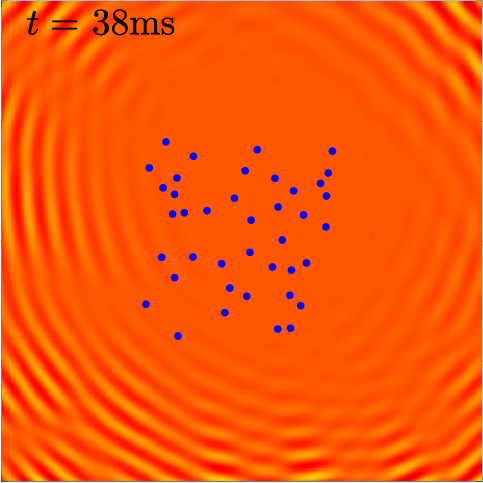}   
		\end{tabular}			
		\caption{Time domain simulation of a Gaussian pulse centered at frequency $\frac{\omega_0}{2\pi} =850$ Hz with profile given in fig. \ref{fig09}. Top: exact solution based on MST. Bottom: Born approximation.}%
		\label{fig12}%
	\end{center}
\end{figure}

Figure \ref{fig12} shows the time-domain simulation of the plate's response to the incident pulse wave, generated by the action of the point force. Snapshots at different moments of the simulation are represented: in the top two rows, the exact solution obtained with multiple scattering, and in the bottom two rows, the solution based on the Born approximation, with the frequency-dependent scattered wavefield expression given by Eq.  \eqref{eq015}. The first row of each simulation roughly covers the time range from when the incident wave reaches the first resonator until it passes through the last one, that is, in the interval \(20 \leq t \leq 28\) ms. In this time window, the greatest scattering intensity occurs, which directly results from the effect of the incident wave on the scatterers. This is precisely the zone of influence of the Born approximation, and as expected, the results between both models are very similar, despite the frequency spectrum of the temporal signal covers a broad band, including the region where the spectral radius is greater than one, as shown in Fig. \ref{fig10} The iterative method developed in Eqs. \eqref{eq087} is strictly divergent outside the so-called weak scattering region associated to \(\rho(\mathbf{g}) < 1\), meaning that the matrix cannot be inverted using the Neumann series in the region where \(\rho(\mathbf{g}) > 1\).  However, the results obtained provide a rather satisfactory estimation of the plate's response, especially when evaluating the first-order scattering.  As the scattering order increases,  after the wave passes through and due to successive higher-order interactions between scatterers, the proposed approximation solution becomes less precise. It is observed for $t \geq 32$ ms (second row of snapshots) that a higher-order scattered wave field remains between the scatterers in the exact solution, not being able the Born approximation to reproduce such behaviour. On the contrary, we can observe that the field is practically at rest at that point for $t \geq 38$ ms within the region of scatterers, something  expected since the scattered field considered by the Born approximation is directly proportional to the incident field. This phenomenon can clearly be visualized in the complete simulation,   provided in the form of a video in the supplementary material attached to this article. 

\section{Numerical example 3: geometrical distribution comparison}

In this third example, we will emphasize the relevance of the geometric distribution of scatterers and its relation to the spectral radius of matrix $\bm{\mathcal{G}}$, defined in Eq. \eqref{eq022}. In this case, we will consider infinitely rigid attachments between the masses and the plate. Doing $K_\infty \to \infty$ in Eq. \eqref{eq002} we obtain after some math that 
 \begin{equation}
	\tau_\alpha = G_p T_\alpha = \frac{G_p t_\alpha }{1 - G(\mathbf{0}) t_\alpha} = \frac{i(1-i)}{\sqrt{\pi}} \,  \frac{\displaystyle \frac{m_\alpha}{\varrho h} k^2}{\displaystyle 1 -  \frac{m_\alpha}{\varrho h} k^2 }
	\label{eq089}
\end{equation}
Essentially, we have a monotonic increasing function for $\left|\tau_\alpha\right|$ with frequency: proportional to the mass of the scatterer for the long wavelength range and asymptotically independent of the mass for the high-frequency range. Considering all scatterers identical and far-field assumptions it is
\begin{equation}
	\mathbf{g}  \approx \tau_\alpha(k) \, \bm{\mathcal{G}}(k)  
	\label{eq090}
\end{equation}
where the dependence of the plate wavenumber $k$ is highlighted . As we have seen, the $\bm{\mathcal{G}}$ matrix collects the scattering properties of the distribution of points where the resonators are located and their relation with the wavelength. Our hypothesis is that this matrix contains information about the geometric pattern of this distribution and if periodicity exists, it will be reflected in our spectral radius $\rho(\bm{\mathcal{G}})$).

\begin{figure}[h]%
	\begin{center}
		\begin{tabular}{ccccc}
			\includegraphics[width=4.0cm]{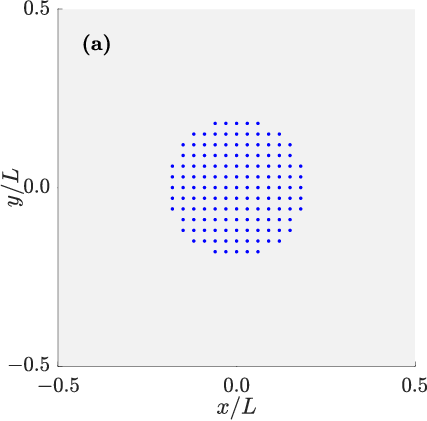}  &
			\includegraphics[width=4.0cm]{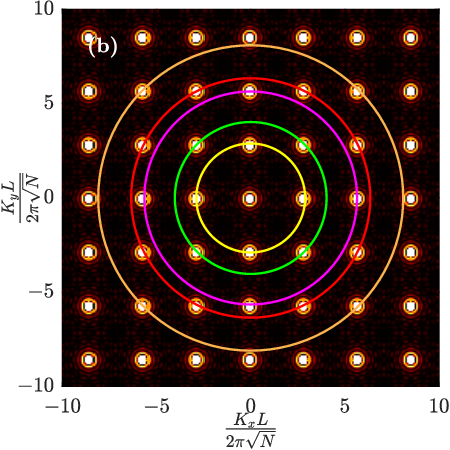}   & 
			\includegraphics[width=4.0cm]{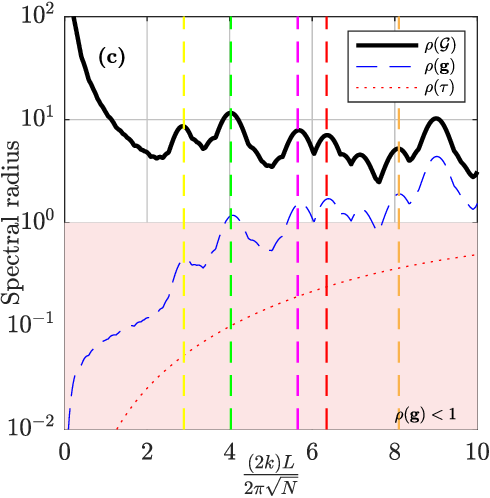}  \\ \\
			\includegraphics[width=4.0cm]{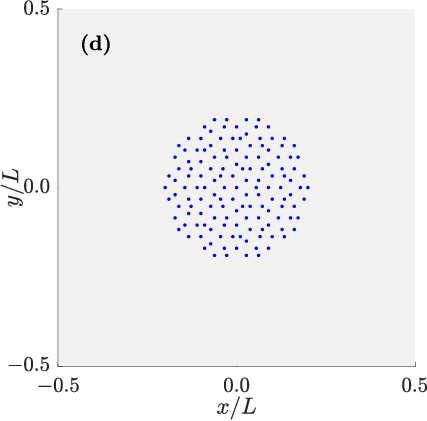}  &
			\includegraphics[width=4.0cm]{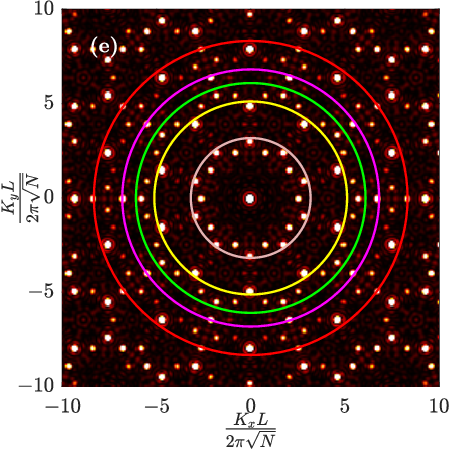}   &
			\includegraphics[width=4.0cm]{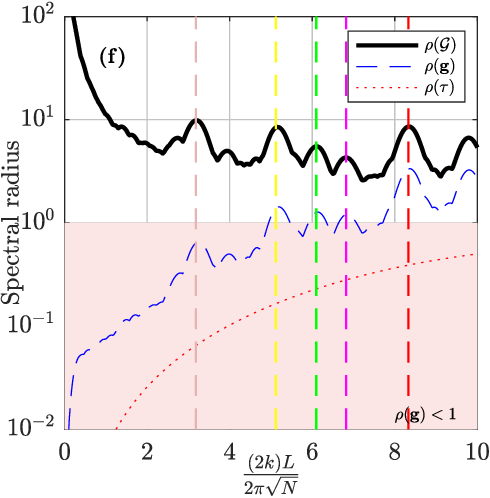}  \\ \\
			\includegraphics[width=4.0cm]{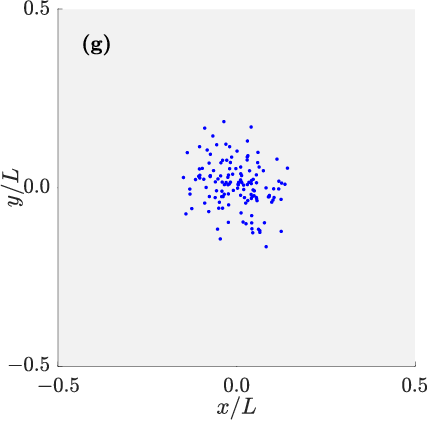}  &
			\includegraphics[width=4.0cm]{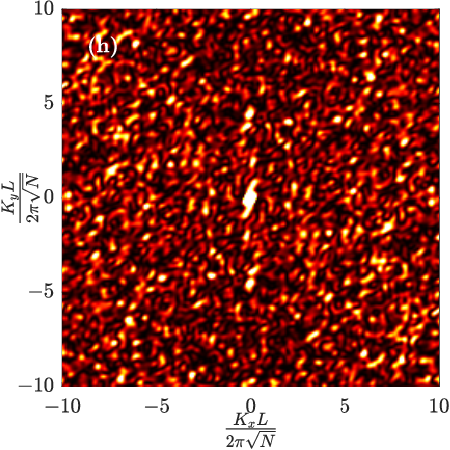}   &
			\includegraphics[width=4.0cm]{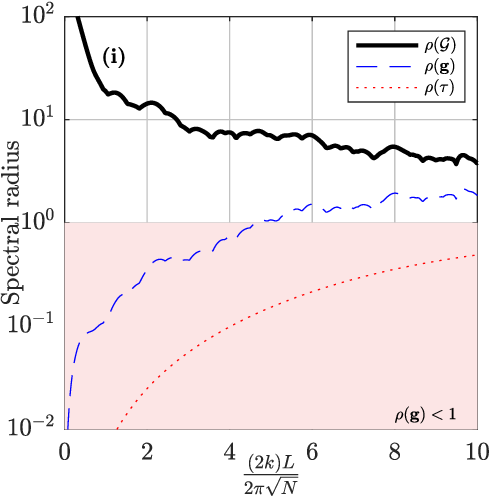}  \\ \\
		\end{tabular}			
		\caption{Three mass clusters with different geometric distributions confined within a circle of 80 cm centered on the 2 mm thick aluminum plate. Point distribution patterns: (a) periodic, (d) quasiperiodic (Penrose crystal), and (g) random. The graphs in the central column show the intensity of the structure factor in reciprocal space for each distribution. The right column reproduces the spectral radius of the scattering matrix \(\mathbf{g}\) and its components, \(\bm{\mathcal{G}}\) and \(\bm{\tau}\), as a function of the plate wavenumber \(k\). The circles centered at the origin of graphs (b) and (e) have radii corresponding to the abscissas of graphs (c) and (f), respectively (same color).}%
		\label{fig13}%
	\end{center}
\end{figure}

Figure \ref{fig13} shows a comparison of the spectral radius of three different geometrical arrangements. An aluminum plate of 2 mm thickness with a cluster of point masses of $m_\alpha = 1$ g  has been considered. Panels (a),(d) and (g) represent three different finite clusters. The three of them have been confined in a 80 cm diameter circular region of the plate. Cluster (a) is made of a periodic square arrangement, cluster (d) is a Penrose crystal and cluster (g) is a random crystal. The number of scatterers in the cluster is also very similar ($137$ in the periodic case, $141$ for the Penrose crystal and $136$ for the random one). To the right of the point distributions, the structure factor associated with the reciprocal lattice (unit cell $L \times L$ = $2 \times 2$ m) and the spectral radii $\rho(\mathbf{g})$ and its components $\rho(\bm{\mathcal{G}})$ and $\rho(\bm{\tau})= \left|\tau_\alpha\right|$, evaluated for each wavenumber $k$. The value of $\rho(\bm{\tau})$ is the same in all three cases, and equal to that predicted by Eq. \eqref{eq089}. However, the value of $\rho(\bm{\mathcal{G}})$ shows pronounced peaks at certain frequencies in the periodic and quasiperiodic case. In these cases, it is found that the half-wavelength associated with the peaks coincides with points of the reciprocal lattice. To highlight this fact, circles have been drawn in plots (b) and (e) with the radii of the peaks found in plots (c) and (f), respectively. {We note that for the periodic case, the peaks of the spectral radius and consequently the circles in the reciprocal space are related to the limits of the several irreducible Brillouin zones due to periodicity.} As expected, a random distribution of points does not contain order or geometric patterns, see plots (g) to (i). The spectral radius  {decreases with frequency} (due to the effect of term $1/\sqrt{k}$, as happens with other clusters) but in this case with no remarkable peaks, except those due to the randomness of the sample. \\

Note that the spectral radius is a quantity derived from the Physics of the plate, while the reciprocal lattice is a purely geometrical property. The link between them makes the spectral radius of the scattering matrix a consistent proposal to evaluate the (i) scattering intensity, (ii) its source and (iii) the strict conditions to apply the Born approximation.

\section{Conclusions}
\label{sec:Conclusions}

In this work, we have developed a theoretical framework for analyzing weak scattering in thin elastic plates with point-like resonators. By employing the Born approximation and the far-field asymptotics of the Green function, we have demonstrated that the wave response can be expressed as a power series expansion, where each term accounts for higher-order scattering interactions. The convergence of this series is directly governed by the spectral properties of the scattering matrix, providing a natural criterion to distinguish between weak and strong scattering regimes.  \\

A key outcome of our formulation is the ability to separate geometric and physical contributions to the scattering process. This separation allows for a systematic evaluation of the conditions under which weak scattering occurs, making it possible to predict when the Born approximation yields accurate results. Through numerical examples, we have validated our theoretical predictions by analyzing periodic and random distributions of resonators. Our results show that weak scattering is strongly influenced by both the resonator properties and their spatial arrangement, with distinct behaviors emerging at Bragg frequencies and near local resonances.  Furthermore, our study highlights the role of the spectral radius of the scattering matrix as a dimensionless measure of scattering intensity. We have shown that this quantity can be approximated as the product of two independent factors: one capturing the global spatial arrangement of the scatterers and the other describing their local impedance relative to the plate. This decomposition provides a clear physical interpretation of scattering phenomena in flexural wave systems and offers a practical tool for designing metamaterials with tailored wave responses.  \\

The implications of our findings extend to various applications in wave control, including vibration isolation, energy harvesting, and the design of stealth materials. The ability to rigorously define weak scattering conditions paves the way for efficient low-order modeling strategies, enabling the optimization of metamaterial structures with minimal computational cost. Future research could explore the extension of this framework to more complex scatterer configurations, non-linear interactions, and dynamic tuning mechanisms for adaptive wave control.

\section*{Acknowledgements}

M.L. and V.R.-G. are grateful for the partial support under Grant No. PID2020-112759GB-I00 funded by MCIN/AEI/10.13039/501100011033. M.L., M. M.-S. and V.R.-G.  are grafeful for the partial support under Grant No. PID2023-146237NB-I00 funded by MICIU/AEI/10.13039/501100011033. M.L. and V.R.-G. acknowledge support from 
Grants CIAICO/2022/052 and CIAICO/2024/318 of the ``Programa para la promoci\'on de la investigaci\'on cient\'ifica, el desarrollo tecnol\'ogico y la innovaci\'on en la Comunitat Valenciana'' funded by Generalitat Valenciana. M.L is grateful for support under the ``Programa de Recualificaci\'on del Sistema Universitario Espa\~nol para 2021-2023'', (funded by ``Instrumento Europeo de Recuperaci\'on (Next Generation EU) en el marco del Plan de Recuperaci\'on, Transformaci\'on y Resiliencia de Espa\~na'', a trav\'es del Ministerio de Universidades. R.V.C acknowledge financial support from the EU H2020FET-proactive project MetaVEH under grant agreement number 952039. 

\section*{Declaration of competing interest}
The authors declare that they have no known competing financial interests or personal relationships that could have appeared to influence the work reported in this paper.

\section*{CRediT authorship contribution statement}
\textbf{ML:} Conceptualisation, Data curation, Formal Analysis, Funding acquisition, Investigation, Methodology, Resources, Software, Validation, Visualization, Writing - original draft, Writing - review \& editing. 
\textbf{MMS:} Conceptualisation, Data curation, Formal Analysis, Investigation, Methodology, Resources, Validation, Visualization, Writing - review \& editing.
\textbf{RC:} Funding acquisition, Investigation, Resources, Supervision, Validation, Visualization, Writing - review \& editing.
\textbf{VRG:} Funding acquisition, Investigation, Resources, Supervision, Validation, Visualization, Writing - review \& editing.

\section*{References}
\bibliographystyle{elsarticle-num} 
\bibliography{bibliography.bib,bibliography2.bib}

\appendix





\end{document}